\documentstyle[aps,epsfig]{revtex}

\begin{document}
\vspace*{0.2in}
\begin{center}
{\LARGE Effect of In-Medium Meson Masses on Nuclear Matter Properties}\\
\vspace*{0.3in}
Abhijit Bhattacharyya\footnote{abhijit@veccal.ernet.in}, \\ 
\centerline{\it  Variable Energy Cyclotron Center, 1/AF Bidhan Nagar,
Calcutta 700 064, India}
Sanjay K. Ghosh\footnote{phys@boseinst.ernet.in} \\
\centerline{ \it Department of Physics, Bose Institute,
93/1, APC Road, Calcutta 700 009, India}
and \\ 
S. C. Phatak
\footnote{phatak@iopb.res.in}\\
\centerline{\it Institute of Physics, Sachivalaya Marg,
Bhubaneswar 751 005, India}
\end{center}

\vspace{0.3cm}
\begin{abstract}
Masses of hadrons (baryons as well as mesons) are modified in the nuclear 
medium because of their interactions. In this paper we investigate the
effect of in-medium meson masses on the properties of the nuclear matter.
The calculations are performed in the Walecka model. We find that with
the inclusion of meson mass modification, the computed equation of state
becomes softer. We also find that the decrease in the nucleon as well
as meson masses in the medium is smaller than that obtained in the
case when the in-medium meson masses are not taken into account. 
\end{abstract} 

PACS numbers: 21.65.+f, 26.60.+c

\vspace{0.3cm}

\section{Introduction}

Walecka model\cite{a,b} and its extensions, which include nonlinear meson 
interactions\cite{c} and derivative coupling\cite{d}, have been used 
extensively in the investigation of nuclear systems for quite some time
now. One of the attractive features of these models is that these are
relativistic models based on meson-nucleon interactions. The saturation
properties of the nuclear systems is in-built in these models through
scalar ($\sigma$) and vector ($\omega$ and $\rho$) mesons interacting
with the nucleons. The scalar meson generates long range attraction and
the vector mesons give rise to short range repulsion. Furthermore, the 
spin-orbit interaction which is essential in nuclear physics arises in
a very natural way in these models. A number of nuclear structure
calculations based on the extensions of the Walecka model\cite{e}
have been performed. These calculations have succeeded in explaining 
the properties of stable nuclei across the periodic table as well as 
those of nuclei far away from the
stability line. In a nut-shell, the calculations in the Walecka model
are performed in a mean field approximation where one solves the nucleon
and meson (classical) field equations self-consistently. Thus, the
calculations are done in Hartree approximation and Fock (exchange) terms
are not included. Essentially, this amounts to saying that the meson
fields develop nonzero vacuum expectation values in the presence
of nucleons and these mean fields, in turn, generate scalar and vector 
potentials for the nucleons. The effective (in-medium) nucleon mass
($m_n^*$) is then the sum of free nucleon mass and the scalar potential.

The success of the Walecka-type models in explaining the nuclear
properties has led to the application of these models to other areas.
This essentially amounts to extrapolating the applicability of the
model to large isospin, nuclear densities much larger than the normal
nuclear matter density ($\rho_{NM}$) and nonzero temperatures. For
example, the model has been used to compute the equation of state of
neutron matter which is used in neutron star calculations\cite{f}.
It has also been employed in the calculations of nuclear equation of state
at densities higher than $\rho_{NM}$ \cite{g}. 
Recently calculations of in-medium
meson masses at high densities and temperatures in Walecka 
model have also been reported\cite{h,h1}. These calculations were motivated 
by the
conjecture that the properties of hadrons may be modified in nuclear
medium\cite{i} and this may have significant consequences on the
reactions occuring in the nuclear medium. These calculations show that
$\rho$ and $\omega$  meson masses are significantly altered in the
nuclear medium. For example, $\omega$ mass may be as low as $630 MeV$ 
even at $\rho_{NM}$ which is about 0.8 times the free $\omega$ mass 
$m_{\omega}$.

The Walecka model was initially introduced as an effective model for the
calculation of properties of nuclei. Thus the scalar and vector mesons
were effective fields introduced to account for the main features of the
N-N interaction. It is true that the pion, which generates the long
range NN interaction is not included. But it is argued that in the mean 
field approximation for spin and/or isopin zero systems one pion
exchange does not contribute and the two pion exchange is incorporated
through $\sigma$ meson exchange. This argument breaks down at high
temperatures where pionic excitations should dominate. If one takes this
argument seriously, in the spirit of effective interactions of nuclear
structure or the Fermi theory of weak interaction, then it is not
meaningful to compute the in-medium meson masses in Walecka model. That
is, since the parameters of the model (meson masses and the coupling
constants) are fitted to satisfy the nuclear properties (binding
energies, nuclear incompressibility, symmetry energy etc.), it does not
make sense to compute the in medium meson masses again. In other words,
the fitted meson masses are in medium masses and therefore should not be
computed again. 

The argument given in the preceding paragraph is reasonable if the
variation of the effective masses as a function of nuclear density or
temperature is weak. However, the calculations \cite{h} show that the
effective masses change quite a bit as a function of density. This means
that the fitted value of the meson masses at $\rho_{NM}$  will change
significantly at higher densities and the computation of the nuclear
matter properties at these densities will not be reliable. 

Clearly, a better procedure would be to use the in-medium meson masses
in Walecka model calculations particularly if the results are
sensitive to the meson masses. We do believe that such self-consistent
calculations would yield results different from the standard Walecka 
model calculations. To illustrate this point, let us consider nucleons 
interacting with scalar mesons of mass $m$. The nucleon-nucleon
interaction potential generated by the scalar meson exchange is then
$V(r)=g^2{{  e^{-mr}}\over {4\pi r}}$ where $g$ is the meson-nucleon
coupling constant (see for details \cite{j}). If we consider the meson mass 
in the medium is
modified to $m_{eff}$, the nucleon-nucleon potential in the medium is
obtained by replacing $m$ by $m_{eff}$. Clearly, this change will affect
the computation of energy/nucleon, effective nucleon mass etc. In
particular, the decrease in the meson mass will lead to increase in the
range of the potential and decrease in its strength. Further more, since
the meson mass in the medium depends on the nuclear density, the strength
and the range of the potential will be density-dependent. Thus, one expects
that nuclear matter properties computed using this procedure are likely
to differ significantly from the standard Walecka model calculations
particularly at densities larger than $\rho_{NM}$.

The preceeding discussion clearly brings out the problems associated
with the computation of the meson masses in the medium blindly. The
correct procedure should be to compute meson masses as well as nuclear
matter properties self-consistently (upto certain order in perturbation)
and fit the nuclear matter properties by adjusting the parameters of the
Walecka model. Note that this self-consistency is over and above the
self-consistency in the standard Walecka model calculations. We have
done this in the present work. Here we have employed the Walecka model
with $\sigma$ and $\omega$ mesons coupling to nucleons. From our
calculations we find that it is indeed possible to carry out such a
program and fit the nuclear matter properties (binding energy/nucleon at
nuclear matter density). We also find that the change in the nucleon and
meson masses in the nuclear medium are less pronounced and the computed
equation of state is softer. The computed nuclear matter incompressibility
is about 460 MeV in comparison with 610 MeV, the value computed in
standard the Walecka model.

Before we go on to the details of our calculation we would like to note that
our calculations can be repeated for the extensions of the Walecka model
\cite{c,d}. We have not done this here since we want to focus on the effect of 
the variation of meson masses on the nuclear matter properties as a function of
nuclear density. We believe one would obtain qualitatively similar results for
the extensions of Walecka model.

\section{Meson Masses And Nuclear Matter Properties}

The Walecka model Lagrangian, having nucleons interacting with scalar 
$\sigma$-meson and vector $\omega$-meson is given by \cite{b}
\begin{eqnarray}
{\cal L}_W &=& {\bar \psi} (i \gamma_\mu \partial^\mu  - m_n) \psi 
- g_\omega {\bar \psi} 
\gamma_\mu \psi \omega^\mu - {1 \over 4}G_{\mu 
\nu} G^{\mu \nu} + {1 \over 2} m_\omega^2 \omega_\mu \omega^\mu  
\nonumber\\
& & + {1 \over 2}(\partial_\mu \sigma \partial^\mu \sigma - m_\sigma^2 
\sigma^2) +g_\sigma {\bar \psi} \sigma \psi 
\end{eqnarray}
 
In the above equation $\psi$, $\sigma$ and $\omega^\mu$ are, 
respectively, the nucleon, the scalar and the vector meson fields; 
$m_n, m_\sigma$ and $m_\omega$ are the corresponding masses; $g_\sigma$ and 
$g_\omega$ are the couplings of the nucleon to scalar and vector 
mesons, respectively and $G_{\mu \nu} = \partial_\mu \omega_\nu - \partial_
\nu \omega_\mu$.  

\begin{figure}[h]
\epsfxsize=13cm
\centerline{\epsfbox{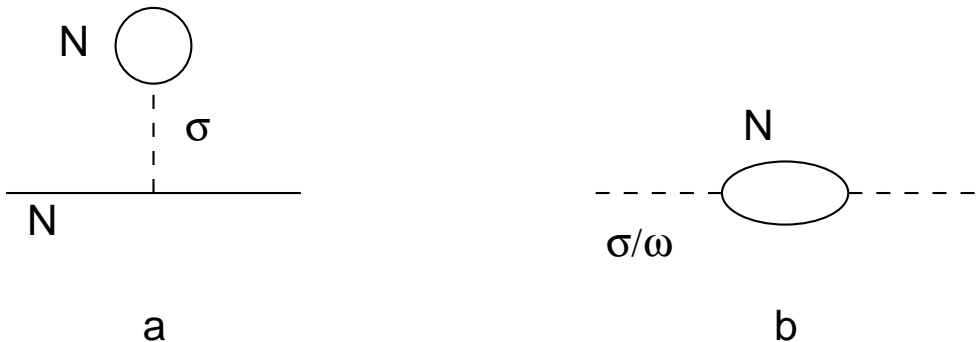}}
\caption{Diagrams contributing to nucleon self energy ( a ) and $\sigma$
and $\omega$ self energies ( b ).}
\end{figure}

In the following discussion we have described the study of the in-medium
 properties of hadrons using the above Lagrangian. The
in medium nucleon mass, the energy density per baryon and the 
$\sigma$ and $\omega$ masses are calculated self-consistently. The  
coupling constants are determined by fiting the 
nuclear matter saturation density to be $\rho_{NM} = 0.15 fm^{-3}$, 
the corresponding binding energy being $-15.75 MeV$ per baryon. 

The effective nucleon mass in the Relativistic Hatree Approximation
(RHA) can be calculated from figure 1a. This is given by \cite{b} 
\begin{eqnarray}
m_n^* &=& m_n - g_\sigma \sigma_0 \nonumber\\
&=& m_n - {{ig_\sigma^2} \over {m_\sigma^{*2}}} \int{{d^4p} \over {(2\pi)^4}}
Tr. G(p)
\end{eqnarray}
where $G$ is the in-medium baryon propagator which is 
given by
\begin{equation}
G(p) = (\gamma^\mu p_\mu + m_n^*) \left[{1 \over {p^2 - m_n^{*2} +
i\epsilon}} + 2i\pi \delta(p^2 - m_n^{*2}) \theta(p_0 - k_F) \right] 
\end{equation}
So we get 
\begin{eqnarray}
m_n^* & = & m_n - {{g_\sigma^2} \over {\pi^2 m_\sigma^{*2}}}
m_n^*\left[k_F E_F  
+m_n^{*2}ln \left({{k_F+E_F} \over m_n^*}\right) \right] 
+{{g_\sigma^2} \over {\pi^2 m_\sigma^{*2}}} \left[m_n^{*3} 
ln\left({m_n^* \over m_n}\right) \right . \nonumber \\
 & & \left . + {1 \over 4} (m_n^{*3} - m_n^3) + {7 \over 4} m_n^2(m_n - m_n^*)  
- {13 \over 4} m_n(m_n-m_n^*)^2 + {25 \over 12} (m_n-m_n^*)^3 \right] 
\end{eqnarray}

The effective mass of a meson can be calculated from the solution of the 
Dyson-Schwinger equation which relates the free and the full propagator 
as follows :
\begin{equation}
D(p) = D_0(p) + D_0(p) \Pi D(p)
\end{equation}
where $D_0(p)$ is the free Green's function, $D(p)$ is the full Green's 
function and $\Pi$ is the polarisation function (at zero three momentum). 
The pole of the full 
propagator then gives the effective mass of the meson. 

We calculate the meson masses in Random Phase Approximation (RPA). The
corresponding diagram is given in figure 1b. 
The effective mass of the $\sigma$ meson is given by \cite{k}
\begin{equation}
m_\sigma^{*2} = m_\sigma^2 + \Pi_\sigma
\end{equation}
where 
\begin{eqnarray}
\Pi_\sigma & = & -i g_\sigma^2 \int {{d^4p} \over {(2\pi)^4}} G(p+q) G(p) 
\nonumber\\
& = & - {{8g_\sigma^2} \over \pi^2} \int_0^{k_F} 
{{p^4 dp} \over E_p^*}  {1 \over {\left(m_\sigma^{*2} -4E_p^{*2}\right)}} 
+{{g_\sigma^2} \over \pi^2} \left[
{3 \over 2} \left(m_n^{*2} - m_n^2 \right) \int_0^1 dx \; {\rm ln}  \left(
1 - {m_\sigma^2 \over m^2} x (1-x) \right)  
\right . \nonumber \\
& & \left . +{3 \over 2} \int_0^1 dx \left(m_n^{*2}-m_\sigma^{*2}x(1-x)\right)
{\rm ln} \left \{ {{ m_n^{*2}-m_\sigma^{*2}x(1-x)} \over {m_n^2-m_\sigma^2
x(1-x)}} \right \}
-  {{m_\sigma^2 - m_\sigma^{*2}} \over 4} \right ]
\end{eqnarray}

The polarisation function for the $\sigma$-meson has two parts. First
part is the in-medium contribution and the second part is the vacuum
contribution. In calculating the vacuum part we have renormalised the
contribution according to ref. \cite{l}. 

For the omega meson we have 
\begin{equation}
m_\omega^{*2} = m_\omega^2 + \Pi_{\omega T}
\end{equation}
where $\Pi_{\omega T}$ is the transverse part of the 
polarisation tensor $\Pi_{\mu \nu}$ (in the static limit the transverse
and the longitudinal components are the same) which is given by
\begin{equation}
\Pi_{\mu \nu} = -i g_\omega^2 \int {{d^4p} \over {(2\pi)^4}} \gamma_\mu 
G(p+q) \gamma_\nu G(p) 
\end{equation}

The transverse part is given by \cite{h}
\begin{equation}
\Pi_{\omega T} = -{{8g_\omega^2} \over {3\pi^2}} \int_0^{k_F} 
{{p^2 dp} \over E_p^*} {{(2p^2 + 3m_n^{*2})} \over 
{\left(m_\omega^{*2} -4E_p^{*2}\right)}} 
-{{g_\omega^2 m_\omega^{*2}} \over \pi^2} 
\int_0^1 dx ln{{ m_n^{*2}-m_\omega^{*2}x(1-x)} \over {m_n^2-m_\omega^2x(1-x)}} 
\end{equation}

The binding energy per baryon, calculated in RHA, is given by,
\begin{equation}
\epsilon = {{\cal E} \over {\rho_{NM}}} -m_N
\end{equation}
where ${\cal E}$ is the energy density which is $\langle T^{00}
\rangle$ and $T^{\mu \nu}$ is the energy momentum tensor. The energy
density in the RHA is given by 
\begin{eqnarray}
{\cal E}& = &{{g_\omega^2} \over{2m_\omega^{*2}}}\rho_{NM}^2 + {{m_\sigma^{*2}}
\over {2g_\sigma^2}}(m_n-m_n^*)^2 + {2 \over \pi^2} 
\int^{k_F}_0 dp p^2 {\sqrt{(p^2 + m_n^{*2})}} -{1 \over {4\pi^2}}
 \left[m_n^{*4} \;{\rm ln} \left({m_n^* \over m_n}\right) \right . 
\nonumber \\
& & \left .+m_n^3(m_n-m_n^*) - {7 \over 2} m_n^2(m_n-m_n^*)^2 
 + {13 \over 3} m_n(m_n-m_n^*)^3 - {25 \over 12} (m_n-m_n^*)^4 \right] 
\end{eqnarray}

We now solve the above equations self-consistently. The parameter values
that we get, from the nuclear matter saturation properties, are given
by $g_\sigma = 6.716$ and $g_\omega = 7.64$. In case we do not consider the
in-medium meson masses the corresponding values are $8.69$ and $9.9$
respectively. The results are shown in
figures 2-3. In figure 2 we have plotted the masses of nucleon, scalar meson
and vector meson as a function of density. In order to make a comparison,
both the cases {\it i.e.} with and without self-consistency have been
plotted. In figure 3 the energy per baryon has been plotted for both cases.

\section{Results and Discussion}

We now come to the discussion of the results. Figure 2 displays the plots of
hadron masses as a function of nuclear density. The dashed lines correspond 
to the standard Walecka model calculation where the modification of meson
masses are not included in the calculation of nuclear matter properties and 
the continuous lines correspond to our (self-consistent) calculation. We 
find that in a self-consistent calculation the change in the masses 
is somewhat less dramatic. In particular, the change in the  
in-medium $\sigma$ and $\omega$ 
masses is less than 
15\% in a self-consistent calculation whereas the maximum change is more than
 20\% for
the standard Walecka model. Hence at nuclear matter density the $\sigma$ and
$\omega$ masses are $471 MeV$ and $686 MeV$ respectively for the 
self-consistent case, whereas the corresponding values for the case 
where there is no
self-consistency are $445 MeV$ and $632 MeV$ respectively. This in terms of the
$N-N$ interaction potential implies that the potential is more attarctive for
the self-consistent case. As a result the nuclear matter will be more 
compressible {\it i.e.} the incompressibility will be lower. Such changes, 
in general, will have bearing on the binding energy as well as on the 
equation of state of the system.  The behavior of the nucleon mass 
appears to be similar for both the cases but the effective nucleon 
mass is slightly larger in a self-consistent calculation ( 0.79 $m_N$ {\it vs } 
0.73 $m_N$ at $\rho_{NM}$ ). From the behavior of the masses one
can conclude that in order to obtain a reliable estimate of
masses of hadrons in nuclear medium it is essential to do a self-consistent
calculation. 

\begin{figure}[h]
\epsfysize=14cm
\centerline{\epsfbox{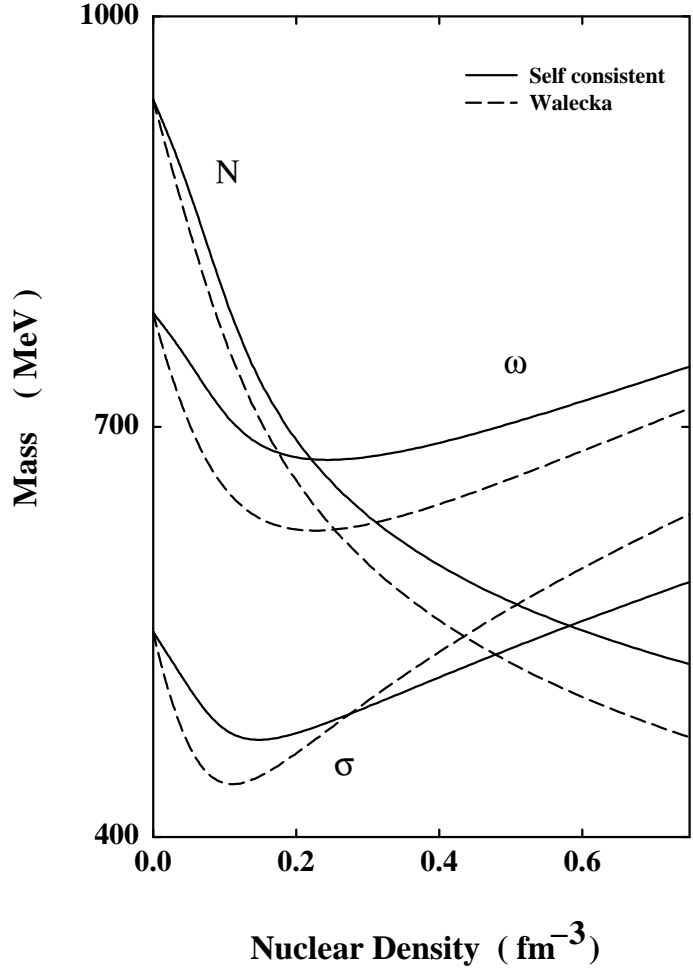}}
\caption{Nucleon, $\omega$ and $\sigma$ masses
effective masses. Continuous and dashed lines are for with
and without self-consistancy.}
\end{figure}

Figure 3 shows a plot of the binding energy/nucleon ( $B/A$ ) as a function 
of the
nuclear density. Both the curves have a minimum at the nuclear matter density, 
simply because the binding energy at the saturation density is fitted
($-15.75 MeV$ at $0.15 fm^{-3}$). One should, however, note that the second 
derivative 
of $B/A$, which is related to the nuclear incompressibility, is smaller for the
self-consistent calculation. The computed incompressibility for the 
self-consistent
calculation is 460 MeV in comparison with the standard Walecka model result
of 610 MeV. In other words, the self-consistent calculation drives the  
incompressibility towards lower values. The binding energy per baryon is much
lower at higher densities for the self-consistent case. This implies that the
equation of state obtained in the self-consistent calculation is softer.        This is an important result in the context of
neutron star physics. In the standard Walecka model the equation of state is
stiffer resulting a value of the neutron star mass which is higher than the
experimental bound. Our results indicate that a self-consistent calculation
may yield a better estimate for the neutron star masses. 

So, to conclude, we have calculated the nuclear matter properties as well
as the in-medium meson masses self-consistently by including the change
in the meson masses due to medium effect in the calculation of nuclear
matter properties. Our study gives rise to two important conclusions.
First of all, we find that the change in the meson masses due to the
effect of the medium is smaller in a self-consistent calculation. This
means that a calculation which does not impose the self-consistency over
estimates the meson masses. The second result is that the
self-consistent calculation yields a softer equation of state and lower
values of incompressibility. We believe that this qualitative behaviour
will persist for asymmetric nuclear matter and will have a bearing on
neutron star properties.  

\begin{figure}[h]
\epsfysize=14cm
\centerline{\epsfbox{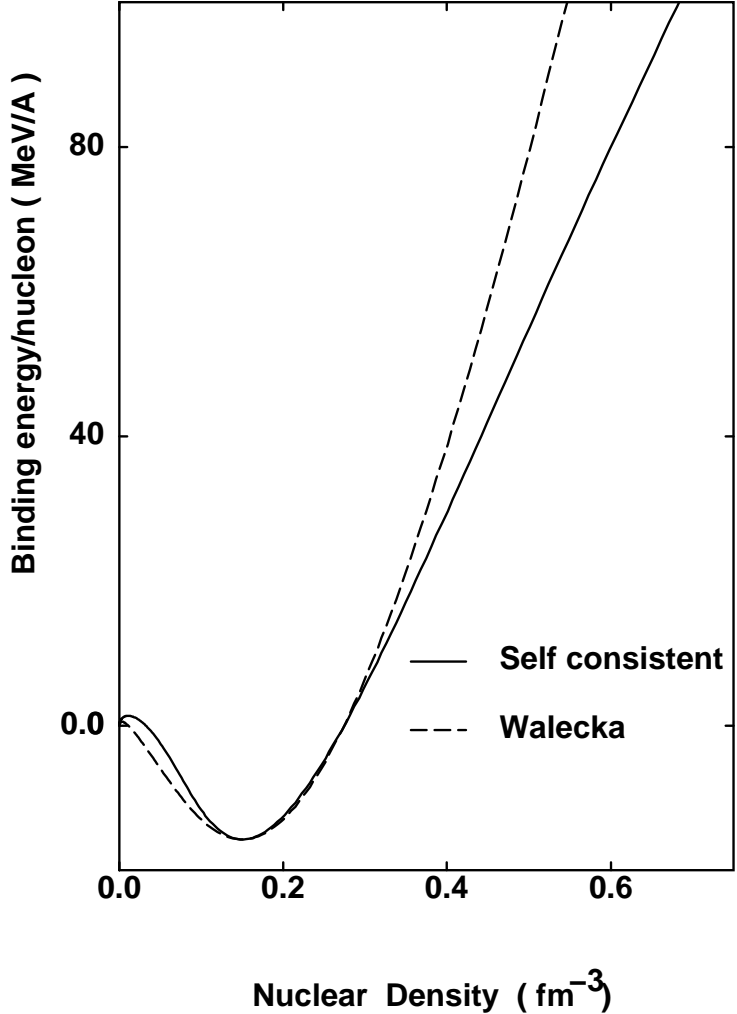}}
\caption{Binding energy/nucleon ( B/A ) as a function of nuclear density.
The continuous and dashed curves are for with and without self-consistancy}
\end{figure}

We would like to emphasize again that, even if the meson fields are
considered as effective fields, the self-consistency described in the 
present work is necessary if one wants to apply the Walecka model at
higher densities. On the other hand if these fields are considered 
as real fields, it is essential to do a self-consistent calculation to
obtain a reliable estimate of meson masses in the medium. It would be
interesting to investigate the behaviour of $\rho$ meson mass in the
medium since it has a bearing on the quark-gluon plasma diagnostics. 
Such a calculation is in progress.

\section{Acknowledgement}
We would like to thank Prof. Sibaji Raha for useful discussions.
AB would like to thank Department of Atomic Energy (Government of India)
and SKG would like to thank CSIR for financial support.

\end{document}